\documentclass[pre,preprint,showpacs,eqsecnum]{revtex4}
\begin{document}
\title{Anomalous diffusion for overdamped particles driven by
cross-correlated white noise sources}
\author{S.~I.~Denisov,$^{1,2}$ A.~N.~Vitrenko,$^{2}$
W. Horsthemke,$^{3}$ and P.~H\"{a}nggi$^{1}$}
\affiliation{$^{1}$Institut f\"{u}r Physik, Universit\"{a}t
Augsburg, Universit\"{a}tsstra{\ss}e 1, D-86135 Augsburg, Germany\\
$^{2}$Sumy State University, 2 Rimsky-Korsakov Street, 40007 Sumy,
Ukraine\\
$^{3}$Department of Chemistry, Southern Methodist University,
Dallas, Texas 75275--0314, USA}

%\date{submitted to Physical Review E: \today}

\begin{abstract}
We study the statistical properties of overdamped particles driven
by two cross-correlated multiplicative Gaussian white noises in a
time-dependent environment. Using the Langevin and Fokker-Planck
approaches, we derive the exact probability distribution function
for the particle positions, calculate its moments and find their
corresponding long-time, asymptotic behaviors. The generally
anomalous diffusive regimes of the particles are classified, and
their dependence on the friction coefficient and the characteristics
of the noises is analyzed in detail. The asymptotic predictions are
confirmed by exact solutions for two examples.
\end{abstract}
\pacs{05.40.-a, 05.10.Gg, 02.50.-r}

\maketitle

\section{Introduction}

Diffusive behavior is intrinsic to many physical, chemical,
biological, economical and other systems \cite{V-K,HL,G}. One of the
most rigorous approaches for its description that takes into account
the dynamical origin of diffusive motion is based on the Langevin
equation, i.e., the stochastic equation of motion. For a large
variety of systems that are stochastically equivalent to free
Brownian particles, this equation has the simplest form, namely a
constant friction coefficient and additive Gaussian white noise.
Such systems exhibit normal diffusion, i.e., the long-time
asymptotics of the dispersion of the state parameter are
proportional to time.

However, plenty of systems exhibit anomalous diffusion (for a
review, see Refs.~\cite{BG,AH,Z,MeKa}). In general, there exist
several approaches  to describe this phenomenon. A first one is
based on fractional calculus and involves fractional diffusion
equations \cite{SW,H,MK}, fractional Fokker-Planck equations
\cite{F,C,MBK1,B} or fractional Langevin equations
\cite{KR,L,BBT}. Yet another other relies on generalized ordinary
Langevin equations. The latter is very attractive for the
treatment of diffusive behavior, and it is especially informative
if those equations can be solved exactly \cite{Hanggi81,IP}. This
approach has been employed to study anomalous diffusion for a
variety of systems, including systems described by the generalized
Langevin equation with a friction memory kernel
\cite{MRNJ,PWM,WT,Morgado,Bao,BHZ,Mokshin}, free undamped
\cite{M,Hein,MW} and damped particles \cite{DH1} driven by
additive noise, overdamped particles driven by one \cite{DH2} and
two \cite{V} multiplicative colored noises, and overdamped
particles with time-dependent drift driven by additive Gaussian
white noise \cite{LM}.

The Langevin approach is also an efficient  tool to study various
effects in systems driven by correlated noises. The correlation
can exist, for example, if the two noises possess a common origin
\cite{FT,MHW}, as in the case of bistable systems where
fluctuations of some model parameters are equivalent to the action
of correlated additive and multiplicative noises \cite{MHW}. Such
noises play an important role in the description of stochastic
resonance \cite{GHJM,H2,LH}, non-equilibrium phase transitions
\cite{JL,DVH}, and noise-induced transport \cite{CW,LLH}. We
expect that correlated noise will impact significantly the
anomalous diffusive behavior. To investigate this problem in the
general case of a time-dependent environment, we consider the
long-time behavior of an overdamped particle governed by the
Langevin equation
\begin{equation}
    \lambda(t)\dot{x}(t)=\sum_{i=1}^{2}g_{i}\textbf{(}x(t)
    \textbf{)}\xi_{i}(t)\quad [x(0)=x_{0}],
    \label{eq:Lang_intr}
\end{equation}
where $x(t)$ is the particle position, $\lambda(t)$ denotes an
arbitrary, positive-valued time-dependent friction coefficient,
and $g_{i}(x)$ are deterministic functions that characterize the
state-dependent action of Gaussian noises $\xi_{i}(t)$. We suppose
that all parameters and variables in Eq.~(\ref{eq:Lang_intr}) are
dimensionless and that the noises are white with $\langle
\xi_{i}(t) \rangle = 0$ and
\begin{equation}
    \langle\xi_{i}(t)\xi_{j}(t')\rangle =
    2\Delta_{ij}\delta(t-t').
    \label{eq:corr_funcs}
\end{equation}
Here $\langle\ldots\rangle$ denotes averaging with respect to the
noises $\xi_{i}(t)$, $\Delta_{11} \equiv \Delta_{1} (\geq0)$ and
$\Delta_{22} \equiv \Delta_{2}(\geq0)$ are the intensities of the
noises $\xi_{1}(t)$ and $\xi_{2}(t)$, respectively, $\Delta_{12} =
\Delta_{21} \equiv r\sqrt{\Delta_{1} \Delta_{2}}$, $r$ is the
parameter characterizing the cross-correlation of the noises, $|r|
\leq 1$, and $\delta(t)$ is the Dirac delta-function. For
generality, we assume that the noises $\xi_{i}(t)$ are external
and so the time-dependent friction and stochastic forces in
Eq.~(\ref{eq:Lang_intr}) are not related to the
fluctuation-dissipation theorem. Hence, if internal noises are
negligible in comparison with external ones, this equation can be
used, e.g., for the description of Brownian particles in a liquid
with a time-dependent temperature. It is important to note,
however, that Eq.~(\ref{eq:Lang_intr}) has a physical meaning for
internal noises as well. Indeed, since according to the Stokes
formula the friction coefficient depends on the particle radius,
this equation describes in the overdamped regime the Brownian
dynamics of spherical particles with time-dependent radii.
Therefore, Eq.~(\ref{eq:Lang_intr}) with internal noises can be
used, e.g., for the study of stochastic dynamics of nucleating and
reacting particles, see in this context also related works and the
given applications therein which considers instead a fluctuating
friction \cite{luczka1,luczka2}. Equation (\ref{eq:Lang_intr})
does not include inertial effects since in most cases they are
important only for a very short initial time interval and can be
safely neglected for larger time scales.

The paper is organized as follows. In Sec.~II, using the Langevin
approach, we derive the time-dependent probability distribution
function (PDF) of $x(t)$ and calculate its moments. In Sec.~III, we
find the long-time asymptotic behavior for moments and classify the
diffusive regimes of the particles. We obtain some exact solutions
in Sec.~IV. In the same section we compare the exact and asymptotic
results. We summarize our results in Sec.~IV. Finally, we derive the
PDF by the Fokker-Planck method in the Appendix.

\section{Probability distribution
function and its moments}

According to the results of Refs.~\cite{FT,MHW,WCK}, the two-noise
Langevin equation (\ref{eq:Lang_intr}) is statistically equivalent
to a one-noise Langevin equation. Its form depends on how
Eq.~(\ref{eq:Lang_intr}) is interpreted \cite{DVH}. In the case of
the Stratonovich interpretation \cite{S,PR} we obtain
\begin{equation}
    \lambda(t)\dot{x}(t)=G\textbf{(}x(t)\textbf{)}\zeta(t)
    \quad [x(0)=x_{0}],
    \label{eq:Lang_1}
\end{equation}
where
\begin{equation}
    G(x)=\sqrt{\sum_{i=1}^{2}\sum_{j=1}^{2}
    \Delta_{ij}g_{i}(x)g_{j}(x)}
    \label{eq:Gx}
\end{equation}
and $\zeta(t)$ is  Gaussian white noise with zero mean and
correlation function
\begin{equation}
    \langle\zeta(t)\zeta(t')\rangle=2\delta(t-t').
    \label{eq:corr_func}
\end{equation}
In general, the function $G(x)$ will be nonnegative. Here, we
consider the slightly more stringent case that $G(x) > 0$ for all
$x$, and consequently $x(t) \in (-\infty,\infty)$.

In order to find the PDF of $x(t)$, we use an explicit expression
for $x(t)$ that follows from Eq.~(\ref{eq:Lang_1}). (An alternative
derivation of the PDF via the solution of the corresponding
Fokker-Planck equation is presented in the Appendix.) The Wang and
Zakai theorem \cite{PR,WZ} states that the conventional rules of
calculus are applicable to Langevin equations if they are
interpreted as Stratonovich stochastic equations. Accordingly we
solve for $x(t)$ by separating the variables and integrating
Eq.~(\ref{eq:Lang_1}) to obtain
\begin{equation}
    \int_{x_{0}}^{x(t)}\frac{dx'}{G(x')}=\int_{0}^{t}
    \frac{\zeta(t')dt'}{\lambda(t')}.
    \label{eq:int}
\end{equation}
If we define
\begin{equation}
    \Psi(x) - \Psi(x_{0}) =\int_{x_{0}}^{x}\frac{dx'}{G(x')}
    \label{eq:Psi}
\end{equation}
and
\begin{equation}
    w(t)=\int_{0}^{t}\frac{\zeta(t')dt'}{\lambda(t')},
    \label{eq:Wt}
\end{equation}
then Eq.~(\ref{eq:int}) can be rewritten in the form
\begin{equation}
    \Psi(x(t)) = w(t) + u_{0}
    \label{eq:sol1}
\end{equation}
[$u_{0}=\Psi(x_{0})$]. According to the definition (\ref{eq:Psi}),
the function $u=\Psi(x)$ is continuous and monotonically increasing;
consequently, we obtain  $\min u = \Psi(-\infty)$, $\max u =
\Psi(\infty)$. On the other hand, the range of the function $w(t)$
is over all real-valued numbers. Therefore, Eq.~(\ref{eq:sol1}) is
applicable at all times only if $\Psi(-\infty) = -\infty$ and
$\Psi(\infty) = \infty$. We assume that these conditions are met and
take into account that the inverse function $x=\Psi^{-1}(u)$ is
single-valued, i.e. it is also continuous and monotonically
increasing, to obtain the unique solution of Eq.~(\ref{eq:Lang_1})
in the form
\begin{equation}
    x(t)=\Psi^{-1}(w(t)+u_{0}).
    \label{eq:sol2}
\end{equation}

The relation (\ref{eq:sol2}) shows that the statistical properties
of $x(t)$ are defined by the Gaussian process $w(t)$, which is fully
characterized by its zero mean and the correlation function
\begin{equation}
    \langle w(t)w(t')\rangle=2\int_{0}^{\min(t,t')}\frac{d\tau}
    {\lambda ^{2}(\tau)}.
    \label{eq:corr_func_W}
\end{equation}
Let $P_{x}(x,t)$ and $P_{w}(w,t)$ be the univariate PDFs that
$x(t)=x$ and $w(t)=w$, respectively. Since a one-to-one
correspondence exists between $x(t)$ and $w(t)$, the condition
$P_{x}(x,t)|dx|= P_{w}(w,t)|dw|$ must hold. Taking into account that
$P_{w}(w,t) = [\sqrt{2\pi}\, \sigma(t)] ^{-1}\exp[-w^{2}/
2\sigma^{2}(t)]$, where
\begin{equation}
    \sigma^{2}(t) = 2\int_{0}^{t}\frac{d\tau}{\lambda ^{2}(\tau)}
    \label{eq:sigma}
\end{equation}
is the dispersion of $w(t)$, and using the relation  that $dx/dw =
G(x)$, which follows from Eqs.~(\ref{eq:Psi}) and (\ref{eq:sol1}),
we obtain for the desired PDF:
\begin{equation}
    P_{x}(x,t)=\frac{1}{\sqrt{2\pi}G(x)\sigma(t)}\exp\!\bigg
    [-\frac{(\Psi(x)-u_{0})^2}{2\sigma^{2}(t)}\bigg]\!.
    \label{eq:Px}
\end{equation}

This distribution function is properly normalized, na\-mely,
$\int_{-\infty}^{\infty} P_{x}(x,t)dx = 1$, and its moments, defined
as
\begin{equation}
    \langle x^{n}(t)\rangle = \int_{-\infty}^{\infty}dx\,
    x^nP_{x}(x,t)
    \label{eq:def_mom}
\end{equation}
($n = 1,2,\ldots$), contain valuable information about the
stochastic system dynamics. Substituting Eq.~(\ref{eq:Px}) into
Eq.~(\ref{eq:def_mom}) and introducing the new variable $y =
(\Psi(x)-u_{0})/\sigma(t)$, we reduce Eq.~(\ref{eq:def_mom}) into
the convenient form
\begin{equation}
    \langle x^{n}(t)\rangle=\frac{1}{\sqrt{2\pi}}
    \int_{-\infty}^{\infty}dy\left[\Psi^{-1}(y\sigma(t) +
    u_{0})\right]^ne^{-y^2/2}.
    \label{eq:mom}
\end{equation}

Depending on the time-dependence of the friction coefficient
$\lambda(t)$, the particles can be localized or delocalized as $t
\to \infty$. If $\sigma(\infty) < \infty$, i.e., if $\lambda(t)$
grows fast enough, then the stationary PDF $P_{x}(x,\infty)$ exists,
and the particles are localized. Brownian particles in a freezing
liquid serve as an example. On the contrary, if $\sigma(t) \to
\infty$ as $t \to \infty$, then the particles are delocalized. We
focus on the latter situation in the following.

\section{Asymptotic behavior for the moments}

To study the long-time behavior of $x(t)$, including various
diffusive regimes, we derive the asymptotic behavior of $\langle
x^{n}(t)\rangle$ for $t \to \infty$. This is a rather complex
problem because the integrand in (\ref{eq:mom}) contains the inverse
function $\Psi^{-1}(u)$ of the integral function $\Psi(x)$. Since
the dispersion $\sigma^{2}(t)$ tends to infinity in the long-time
limit, the inverse function at $y>0$ and $y<0$ can be replaced by
its asymptotics at $u \to \infty$ and $u \to -\infty$, respectively.
To find these asymptotic behaviors, we assume that
\begin{equation}
    G(x) \sim \sqrt{\Delta_{\rho}}\,|x|^{\alpha_{\rho}}
    \quad (x \to \rho\,\infty),
    \label{eq:as_G}
\end{equation}
where $\rho = +$ or $-$, $\Delta_{\rho}$ is the effective white
noise intensity at $x = \rho\, \infty$, and $\alpha_{\rho} \leq 1$
to ensure that $\Psi(\pm\infty) = \pm\infty$. Then for $x \to \rho\,
\infty$, Eq.~(\ref{eq:Psi}) yields
\begin{equation}
    \Psi(x) \sim \frac{\rho}{\sqrt{\Delta_{\rho}}}
    \left\{\begin{array}{ll} \displaystyle
    \ln \frac{|x|}{a_{\rho}}, \quad \alpha_{\rho} = 1,
    \\[14pt]
    \displaystyle \frac{|x|^{1-\alpha_{\rho}}}
    {1-\alpha_{\rho}}, \quad \alpha_{\rho} < 1.
    \end{array}
    \right.
    \label{eq:as_Psi}
\end{equation}
[The normalizing parameter $a_{\rho}$ can be found if the explicit
form of the function $\Psi(x)$ is known (see Sec.~IV).] Finally,
from Eq.~(\ref{eq:as_Psi}) we obtain
\begin{equation}
    \Psi^{-1}(u) \sim \rho a_{\rho}\exp(\sqrt{\Delta_{\rho}}
    \,|u|) \quad (u \to \rho\,\infty)
    \label{eq:as_Psi-1a}
\end{equation}
for $\alpha_{\rho} = 1$, and
\begin{equation}
    \Psi^{-1}(u) \sim \rho[\sqrt{\Delta_{\rho}}\,(1 -
    \alpha_{\rho})|u|]^{\frac{1}{1-\alpha_{\rho}}}
    \quad (u \to \rho\,\infty)
    \label{eq:as_Psi-1b}
\end{equation}for $\alpha_{\rho} < 1$. Next we consider separately
these two cases.

\subsubsection{$\alpha_{\rho} = 1$}

In this case, we replace the inverse function $\Psi^{-1}(y\sigma(t)
+ u_{0})$ by its asymptotic behavior, i.e., $a_{+} \exp
[\sqrt{\Delta_{+}}(y\sigma(t) + u_{0})]$ at $y>0$ and $-a_{-}
\exp[-\sqrt{\Delta_{-}}(y\sigma(t) + u_{0})]$ at $y<0$, and
Eq.~(\ref{eq:mom}) thus yields{\setlength{\arraycolsep}{2pt}
\begin{eqnarray}
    \langle x^{n}(t) \rangle &\sim& \frac{1}{\sqrt{2\pi}} \sum_{\rho}
    \displaystyle(\rho a_{\rho})^{n}\int_{0}^{\infty}dy\,
    \exp(n\sqrt{\Delta_{\rho}}\,(y\sigma(t)
    \nonumber\\[3pt]
    && +\, \rho u_{0}) -y^{2}/2)
    \label{eq:as_x^n_a1}
\end{eqnarray}}as $t \to \infty$. Using the standard integral formula
\cite{PBM}
\begin{equation}
    \int_{0}^{\infty}dy\,e^{-py - y^{2}/2} = \sqrt{\frac{\pi}{2}}\,
    e^{p^{2}/2}\,\textrm{erfc}\bigg(\frac{p}{\sqrt{2}} \bigg),
    \label{eq:int1}
\end{equation}
where $\textrm{erfc}\,(z) = (2/\sqrt{\pi}\,) \int_{z}^{\infty}dt\,
e^{-t^{2}}$ is the complementary error function, and taking into
account the relation $\textrm{erfc}(-\infty) = 2$, we reduce the
asymptotic formula (\ref{eq:as_x^n_a1}) to the form
\begin{equation}
    \langle x^{n}(t) \rangle \sim \sum_{\rho}\displaystyle
    (\rho a_{\rho})^{n}\,\exp(\rho n\sqrt{\Delta_{\rho}}\,
    u_{0} + n^{2}\Delta_{\rho}\sigma^{2}(t)/2).
    \label{eq:as_x^n1}
\end{equation}

It is important to note that, since (\ref{eq:as_Psi-1a}) represents
the leading term of the asymptotic expansion of $\Psi^{-1}(u)$, we
should keep only the largest term in the right-hand side of
(\ref{eq:as_x^n1}). In particular, if $\Delta_{+} > \Delta_{-}$ then
\begin{equation}
    \langle x^{n}(t) \rangle \sim a_{+}^{n}\,\exp(n\sqrt{
    \Delta_{+}}\,u_{0} + n^{2}\Delta_{+}\sigma^{2}(t)/2)
    \label{eq:as_x^n1a}
\end{equation}
($t \to \infty$), i.e., the particles tend to plus infinity, and if
$\Delta_{+} < \Delta_{-}$ then
\begin{equation}
    \langle x^{n}(t) \rangle \sim (-1)^{n}a_{-}^{n}\,\exp
    (-n\sqrt{\Delta_{-}}\,u_{0} + n^{2}\Delta_{-}\sigma^{2}
    (t)/2)
    \label{eq:as_x^n1b}
\end{equation}
($t \to \infty$), i.e., the particles tend to minus infinity. We
emphasize that such a behavior of $x(t)$ results as  a direct
consequence of the multiplicative nature of the noise.

If $\Delta_{+} = \Delta_{-} = \Delta$ then $a_{+} = a_{-} = a$. Both
terms in the right-hand side of (\ref{eq:as_x^n1}) have the same
order, and the asymptotic relation (\ref{eq:as_x^n1}) is reduced to
\begin{equation}
    \langle x^{n}(t) \rangle \sim 2a^{n}\cosh(n\sqrt{\Delta}
    \,u_{0}) \,\exp({n^{2}\Delta\sigma^{2}(t)/2})
    \label{eq:as_x^n1c}
\end{equation}
($t \to \infty$) for even $n$ and to
\begin{equation}
    \langle x^{n}(t) \rangle \sim 2a^{n}\sinh(n\sqrt{\Delta}
    \,u_{0}) \,\exp({n^{2}\Delta\sigma^{2}(t)/2})
    \label{eq:as_x^n1d}
\end{equation}
($t \to \infty$) for odd $n$. The last relation shows that, in
contrast to the previous discussed cases, no systematic growth of
$\langle x(t) \rangle$ occurs for $u_{0} = 0$, i.e., $x_{0} =
\Psi^{-1}(0)$. In other words, in this case the particle position
$x(t)$ exhibits purely diffusive behavior that can be characterized
by the dispersion $\sigma_{x} ^{2}(t) = \langle x^{2}(t)\rangle -
\langle x(t)\rangle^{2}$. Since $\sigma_{x}^{2}(t) \sim 2a^{2}
e^{2\Delta\sigma^{2}(t)}$ as $t \to \infty$, different long-time
regimes of diffusion can exist.

We describe these regimes for the case that $\lambda(t) \sim
lt^{\beta}$ ($t \to \infty$), where $l$ is a positive parameter and
$\beta \leq 1/2$ (the last condition guarantees that $\sigma(\infty)
= \infty$). Using Eq.~(\ref{eq:sigma}), for $\beta < 1/2$ we obtain
\begin{equation}
    \ln \sigma_{x}^{2}(t) \sim \frac{4\Delta}{l^{2}(1-2\beta)}\,
    t^{1-2\beta}
    \label{eq:as_ln_1}
\end{equation}
($t \to \infty$). Accordingly, we call the diffusive behavior a
stretched exponential one if $0 < \beta < 1/2$, exponential if
$\beta = 0$, and a compressed exponential if $\beta < 0$. If $\beta
= 1/2$, then $\sigma^{2}(t) \sim (2/l^{2})\ln(t/\tilde{t})$
[$\tilde{t}$ is some characteristic time scale] and so
\begin{equation}
    \sigma_{x}^{2}(t) \sim 2a^{2}(t/\tilde{t})^
    {\frac{4\Delta}{l^{2}}}
    \label{eq:as_sigma1}
\end{equation}
as $t \to \infty$. This asymptotic formula demonstrates a truly
remarkable feature of this physical system, namely: the character of
its diffusive regime depends on the effective white noise intensity
$\Delta$ which in the general case depends on the correlation
coefficient $r$. Following the conventional terminology, we say that
the system exhibits subdiffusion if $\Delta < l^{2}/4$, normal
diffusion if $\Delta = l^{2}/4$, and superdiffusion if $\Delta >
l^{2}/4$. We expect that these diffusion regimes can exist in the
systems of nucleating particles whose radii grow as $t^{1/2}$.

\subsubsection{$\alpha_{\rho} < 1$}

In this case, using the standard integral formula \cite{PBM}
\begin{equation}
    \int_{0}^{\infty}dy\,y^{\mu}e^{-y^{2}/2} = 2^{\frac{\mu - 1}
    {2}}\,\Gamma\bigg(\frac{\mu+1}{2}\bigg)
    \label{eq:int2}
\end{equation}
[$\mu > -1$, $\Gamma(z) = \int_{0}^{\infty}dx\,x^{z-1}e^{-x}$ is the
gamma function] and the asymptotic relation (\ref{eq:as_Psi-1b}),
Eq.~(\ref{eq:mom}) yields{\setlength{\arraycolsep}{2pt}
\begin{eqnarray}
    \langle x^{n}(t)\rangle &\sim& \frac{1}{2\sqrt{\pi}}
    \sum_{\rho}(\rho1)^{n}\,\Gamma(\eta_{\rho}/2)
    \nonumber\\[3pt]
    && \times [\sqrt{2\Delta_{\rho}}\,(1-\alpha_{\rho})
    \sigma(t)]^{\eta_{\rho}-1}
    \label{eq:as_x^n2}
\end{eqnarray}}($t \to \infty$), where $\eta_{\rho} = 1 +
n/(1-\alpha_{\rho})$. We emphasize that in this asymptotic relation
only the dominant term on the right side should be kept.
Specifically, if $\alpha_{+} > \alpha_{-}$ then
\begin{equation}
    \langle x^{n}(t) \rangle \sim \frac{\Gamma(\eta_{+}/2)}
    {2\sqrt{\pi}}[\sqrt{2\Delta_{+}}\,(1-\alpha_{+})
    \sigma(t)]^{\eta_{+}-1},
    \label{eq:as_x^n2a}
\end{equation}
and if $\alpha_{+} < \alpha_{-}$ then
\begin{equation}
    \langle x^{n}(t) \rangle \sim (-1)^{n}\frac{\Gamma(\eta_{-}/2)}
    {2\sqrt{\pi}}[\sqrt{2\Delta_{-}}\,(1-\alpha_{-})
    \sigma(t)]^{\eta_{-}-1}.
    \label{eq:as_x^n2b}
\end{equation}
Finally, if $\alpha_{\rho} = \alpha$ ($\eta_{\rho} = \eta$) then
both terms have the same order and so{\setlength{\arraycolsep}{2pt}
\begin{eqnarray}
    \langle x^{n}(t)\rangle &\sim& \frac{\Gamma(\eta/2)}
    {2\sqrt{\pi}}[\Delta_{+}^{\frac{\eta-1}{2}} + (-1)^{n}
    \Delta_{-}^{\frac{\eta-1}{2}}]
    \nonumber\\[3pt]
    && \times [\sqrt{2}\,(1-\alpha)\sigma(t)]^{\eta-1}.
    \label{eq:as_x^n2c}
\end{eqnarray}}These relations show that for $\Delta_{+} \neq
\Delta_{-}$ all moments of $x(t)$ diverge as $t \to \infty$.

Next, we assume that $\alpha_{\rho} = \alpha$ and $\Delta_{\rho} =
\Delta$. In this case, using the asymptotic relation
(\ref{eq:as_Psi-1b}), Eq.~(\ref{eq:mom}) for $t \to \infty$ is
reduced to
\begin{equation}
    \langle x^{n}(t) \rangle \sim \frac{\Gamma(\eta/2)}{\sqrt{\pi}}
    [\sqrt{2\Delta}\,(1-\alpha)\sigma(t)]^{\eta-1}
    \label{eq:as_x^n2d}
\end{equation}
if $n$ is even, and to
\begin{equation}
    \langle x^{n}(t) \rangle \sim u_{0}\frac{\eta-1}{\sqrt{2\pi}}\,
    \Gamma\bigg(\frac{\eta-1}{2}\bigg)[\sqrt{2\Delta}\,(1-\alpha)
    ]^{\eta-1}\sigma^{\eta-2}(t)
    \label{eq:as_x^n2e}
\end{equation}
if $n$ is odd. The last relation and the exact formula
(\ref{eq:mom}) confirm the fact that $\langle x(t) \rangle = 0$ for
$u_{0}=0$, i.e., under these conditions the particles display
diffusive behavior. Since in this case
\begin{equation}
    \sigma^{2}_{x}(t) \sim \frac{\Gamma(1/(1-\alpha) - 1/2)}
    {\sqrt{\pi}}[\sqrt{2\Delta}\,(1-\alpha)\sigma(t)
    ]^{\frac{2}{1-\alpha}}
    \label{eq:as_sigma2}
\end{equation}
as $t \to \infty$, its character depends on the exponent $\alpha$
and on the asymptotic behavior of $\sigma(t)$. Specifically, if
$\lambda(t) \propto t^{\beta}$ ($t \to \infty$, $\beta \leq 1/2$)
and so $\sigma^{2}(t) \propto t^{1-2\beta}$ for $\beta < 1/2$ and
$\sigma^{2}(t) \propto \ln(t/\tilde{t})$ for $\beta = 1/2$, then the
particles exhibit superdiffusion if $\alpha > 2\beta$ and $\beta <
1/2$, normal diffusion if $\alpha = 2\beta$ and $\beta < 1/2$,
subdiffusion if $\alpha < 2\beta$ and $\beta < 1/2$, and logarithmic
diffusion $[\sigma^{2}_{x}(t) \propto \ln^{\frac{2} {1-\alpha}}
(t/\tilde{t})]$ if $\beta = 1/2$.

Thus, multiplicativity of noises and time dependency of the
friction coefficient can give rise to anomalous behavior of the
system. Specifically, if $\lambda = \textrm {const}$ then the
conditions $G(x)|_{x \to \pm \infty} \to \infty$ and $G(x)|_{x \to
\pm \infty} \to 0$ are responsible for the fast and slow
diffusion, respectively. On the contrary, if $G(x) = \textrm
{const}$ (white noises are additive) then diffusion is fast if
$\lambda(t)|_{t \to \infty} \to 0$ and is slow if $\lambda(t) |_{t
\to \infty} \to \infty$. Remarkable, the presence of both these
factors can lead to new stochastic phenomena, like the dependence
of the character of anomalous diffusion on the effective white
noise intensity. We note also that the nature of anomalous
diffusion in this system, due the features of the stochastic and
friction forces mentioned above, is quite different from that
observed for random walks with long-tail jump-length and/or
waiting-time distributions.

\section{Specific examples}

To confirm our asymptotic results, we calculate explicit the
expressions for the moments of $x(t)$ for two particular cases.

\subsection{Two additive noises}

As a first example, we consider the case that both white noises are
additive, i.e., $g_{1}(x) = g_{2}(x) = 1$. Then, $G^{2}(x) =
\Delta_{1} + \Delta_{2} + 2r\sqrt{\Delta_{1}\Delta_{2}} \equiv
G^{2}$, $\Psi(x) = x/G$, $\Psi^{-1}(u) = Gu$, $u_{0} = x_{0}/G$, and
Eq.~(\ref{eq:mom}) is reduced to
\begin{equation}
    \langle x^{n}(t)\rangle=\frac{1}{\sqrt{2\pi}}
    \int_{-\infty}^{\infty}dy\left[yG\sigma(t) +
    x_{0}\right]^n e^{-y^2/2}.
    \label{eq:mom1}
\end{equation}
Using in (\ref{eq:mom1}) the binomial formula and taking into
account the relation (\ref{eq:int2}), we obtain after some
calculations
\begin{equation}
    \langle x^{n}(t) \rangle = \frac{x_{0}^{n}}{\sqrt{\pi}}
    \sum_{k=0}^{[n/2]} C_{n}^{2k}\Gamma\bigg(k +
    \frac{1}{2}\bigg)\bigg[\frac{\sqrt{2}\,G}{x_{0}}\,
    \sigma(t)\bigg]^{2k},
    \label{eq:mom2}
\end{equation}
where $[n/2]$ is the integer part of $n/2$ and $C_{n}^{m} =
n!/(n-m)!m!$ is the binomial coefficient.

According to this formula, the leading term of the long-time
asymptotic expansion of $\langle x^{n}(t) \rangle$ is given by the
term in the sum that corresponds to $k = [n/2]$, i.e.,
\begin{equation}
    \langle x^{n}(t) \rangle \sim \frac{1}{\sqrt{\pi}}\,
    \Gamma\bigg(\frac{n+1}{2}\bigg)[\sqrt{2}\,G
    \sigma(t)]^{n}
    \label{eq:as_x^n2f}
\end{equation}
for even $n$ and
\begin{equation}
    \langle x^{n}(t) \rangle \sim \frac{x_{0}}{\sqrt{\pi}}\,n\Gamma
    \bigg(\frac{n}{2}\bigg)[\sqrt{2}\,G \sigma(t)]^{n-1}
    \label{eq:as_x^n2g}
\end{equation}
for odd $n$. Since $\Delta = G^{2}$, $\alpha = 0$ and $\eta = n+1$,
the same asymptotic behavior follows also from the formulas
(\ref{eq:as_x^n2d}) and (\ref{eq:as_x^n2e}), respectively.

\subsection{Additive and multiplicative noises}

As a second example, we consider a linear Langevin equation where
one noise is multiplicative, $g_{1}(x)=x$, and the other one is
additive, i.e., $g_{2}(x)=1$. According to the definition
(\ref{eq:Gx}), in this case
\begin{equation}
    G(x) = \sqrt{\Delta_{1}(x^{2} + 2r\nu x + \nu^{2})}
    \label{eq:G2}
\end{equation}
$(\nu = \sqrt{\Delta_{2}/\Delta_{1}}\,)$ and for $|r| < 1$ we obtain
\begin{equation}
    \Psi(x)=\frac{1}{\sqrt{\Delta_{1}}}\,\mathrm{arcsinh}
    \bigg(\frac{x+r\nu}{\nu\sqrt{1-r^2}}\bigg)
    \label{eq:Psi1}
\end{equation}
and
\begin{equation}
    \Psi^{-1}(u)=\nu\sqrt{1-r^2}\sinh(\sqrt{\Delta_{1}}\,
    u)-r\nu.
    \label{eq:inver_Psi1}
\end{equation}

Then, using Eqs.~(\ref{eq:mom}) and (\ref{eq:inver_Psi1}), the
relation
\begin{equation}
    \sinh(\sqrt{\Delta_{1}}\,u_{0}) = \frac{x_{0} +
    r\nu}{\nu\sqrt{1-r^2}}
    \label{eq:u0_x0}
\end{equation}
that follows from Eq.~(\ref{eq:inver_Psi1}), and the standard
integral formula \cite{PBM}{\setlength{\arraycolsep}{2pt}
\begin{eqnarray}
    \int_{-\infty}^{\infty}\!dy\,e^{-by^{2}+cy}\Bigg[
    \begin{array}{ll}
    \sinh(py) \\
    \cosh(py)
    \end{array}\!
    \Bigg]
    &=& \sqrt{\frac{\pi}
    {b}}\,\exp\!{\bigg(\frac{c^{2}+p^{2}}{4b}\bigg)}
    \nonumber\\ [4pt]
    && \times\,\Bigg[
    \begin{array}{ll}
    \sinh(cp/2b) \\
    \cosh(cp/2b)
    \end{array}\!
    \Bigg]\qquad
    \label{eq:int3}
\end{eqnarray}}($b>0$), we find the first moment
\begin{equation}
    \langle x(t)\rangle = (x_{0} + r\nu)\exp({\Delta_{1}
    \sigma^2(t)/2}) - r\nu,
    \label{eq:first_mom1}
\end{equation}
which shows that $x(t)$ exhibits purely diffusive behavior only if
$r = -x_{0}/\nu$, and the second one{\setlength{\arraycolsep}{2pt}
\begin{eqnarray}
    \langle x^2(t)\rangle &=& [(x_{0} + r\nu)^{2} + \nu^{2}(1 -
    r^{2})/2]\exp[2\Delta_{1}\sigma^2(t)]
    \nonumber\\ [4pt]
    && -\, 2r\nu(x_{0} + r\nu)\exp({\Delta_{1}\sigma^2(t)/2})
    \nonumber\\ [4pt]
    && +\, \nu^{2}(3r^{2}-1)/2 \;.
    \label{eq:sec_mom1}
\end{eqnarray}}For $t \to \infty$, these exact results readily
yield
\begin{equation}
    \langle x(t)\rangle \sim (x_{0} + r\nu)\exp({\Delta_{1}
    \sigma^2(t)/2})
    \label{eq:as_x1}
\end{equation}
and
\begin{equation}
    \langle x^2(t)\rangle \sim [(x_{0} + r\nu)^{2} +
    \nu^{2}(1 - r^{2})/2]\exp[2\Delta_{1}\sigma^2(t)].
    \label{eq:as_x2}
\end{equation}

On the other hand, the same long-time representations for $\langle
x(t)\rangle$ and $\langle x^2(t)\rangle$ follow as well from the
more general asymptotic formulas (\ref{eq:as_x^n1c}) and
(\ref{eq:as_x^n1d}). To verify this,  we compare (\ref{eq:as_G})
with the asymptotic relation $G(x) \sim \sqrt{\Delta_{1}}\,|x|$,
which follows from Eq.~(\ref{eq:G2}) as $|x| \to \infty$; this
yields $\Delta_{\rho} = \Delta = \Delta_{1}$ and $\alpha_{\rho} =
\alpha =1$. Next, since $\mathrm{arcsinh}\,z \sim \rho \ln 2|z|$ as
$z \to \rho\,\infty$, from Eq.~(\ref{eq:Psi1}) we find the
asymptotic formula
\begin{equation}
    \Psi(x) \sim \frac{\rho}{\sqrt{\Delta_{1}}}\,
    \ln\!\bigg(\frac{2|x|}{\nu\sqrt{1-r^2}}\bigg)
    \label{eq:as_Psi1}
\end{equation}
($x \to \rho\,\infty$), which by comparison with (\ref{eq:as_Psi})
yields $a_{\rho} = a = \nu\sqrt{1 - r^{2}}/2$. Finally, substituting
the values for $\Delta$ and $a$ into the formulas
(\ref{eq:as_x^n1d}) [with $n=1$] and (\ref{eq:as_x^n1c}) [with
$n=2$] and using the relation (\ref{eq:u0_x0}), we indeed recover
the asymptotic behavior (\ref{eq:as_x1}) and (\ref{eq:as_x2}).

Thus, our exact results corroborate the general asymptotic formulas
obtained in the previous section.

\section{Conclusions}

We have studied analytically the statistical properties of a special
class of exactly solvable stochastic models represented by an
overdamped particle that is driven by two cross-correlated
multiplicative Gaussian white noises in a time-dependent
environment. As a first step, we have reduced the initial two-noise
Langevin equation that describes the particle dynamics in the
Stratonovich sense to a stochastically equivalent one-noise Langevin
equation. Then, solving the latter equation and the corresponding
Fokker-Planck equation, we have derived the exact probability
distribution function for the particle positions.

To study the long-time behavior of the particle dynamics, we have
calculated the asymptotic behavior for the moments of the
distribution function. Their analysis results in a rich behavior of
differing regimes of anomalous particle diffusion; namely we find
regimes with normal diffusion, subdiffusion, superdiffusion,
exponential diffusion, stretched exponential diffusion, compressed
exponential diffusion, and, as well, logarithmic diffusion. We have
established conditions for these diffusion behaviors to occur and in
some special cases we were able to derive exact formulas for the
moments of order $2$ and higher that confirmed the general
asymptotic results. Also, we have discovered the truly remarkable
feature for systems described by a special class of linear Langevin
equations that their diffusive behavior becomes determined by the
effective white noise intensity.

\section*{ACKNOWLEDGMENTS}

S.I.D. acknowledges the support of the European Union through a
Marie Curie individual fellowship, contract MIF1-CT-2005-007021, and
P.H. the support of the Deutsche Forschungsgemeinschaft, grant HA
1517/13-4.

\appendix*

\section{Alternative derivation of the PDF~(\ref{eq:Px})}

To derive the PDF (\ref{eq:Px}) by the Fokker-Planck me\-thod, we
assume that each Gaussian white noise $\xi_{i}(t)$ is characterized
by its own parameter $\gamma_{i}$ ($0\leq\gamma_{i}\leq 1$) that
determines the points of time at which $g_{i}\textbf{(}x(t)
\textbf{)}$ is evaluated in the corresponding integral sum. Then the
Fokker-Planck equation associated with the Langevin equation
(\ref{eq:Lang_intr}) takes the form \cite{DVH}
\begin{equation}
    \lambda^2(t)\frac{\partial}{\partial t}P_{x}(x,t)=
    -\frac{\partial }{\partial x}h(x)P_{x}(x,t)
    +\frac{\partial^{2}}{\partial x^{2}}d(x)P_{x}(x,t)
    \label{eq:FP}
\end{equation}
[$P_{x}(x,0)=\delta(x-x_{0})$], where
\begin{equation}
    h(x) = 2\sum_{i=1}^{2}\sum_{j=1}^{2}\gamma_{i}
    \Delta_{ij}g'_{i}(x)g_{j}(x)
    \label{eq:def_h}
\end{equation}
is the drift coefficient,
\begin{equation}
    d(x) = \sum_{i=1}^{2}\sum_{j=1}^{2}\Delta_{ij}g_{i}(x)
    g_{j}(x),
    \label{eq:def_d}
\end{equation}
is the diffusion coefficient, and the prime denotes the derivative
with respect to the argument of the function.

In order to solve Eq.~(\ref{eq:FP}), we introduce a new variable $u
= U(x)$, where the function $U(x)$ remains to be defined. If the
functions $U(x)$ and $U^{-1}(u)$ are single-valued and $U'(x)>0$,
then
\begin{equation}
    P_{x}(x,t) = P_{u}(u,t)U'(x).
    \label{eq:rel}
\end{equation}
According to Eq.~(\ref{eq:FP}), the PDF $P_{u}(u,t)$ of the random
process $u(t) = U\textbf{(} x(t)\textbf{)}$ satisfies the
Fokker-Planck equation whose drift and diffusion coefficients (as
functions of the old variable $x$) are given by
\begin{equation}
    \tilde{h}(x)=h(x)U'(x)+d(x)U''(x), \quad
    \tilde{d}(x)=d(x)[U'(x)]^{2}.
    \label{eq:coeff2}
\end{equation}
If we chose $\tilde{d}(x) = 1$, then $U'(x) = 1/\sqrt{d(x)}$ and
\begin{equation}
    \tilde{h}(x) = \frac{2h(x)-d'(x)}{2\sqrt{d(x)}}.
    \label{eq:tilde_h}
\end{equation}

Remarkably, if $\gamma_{i}=1/2$, i.e., if the Stratonovich
interpretation of the Langevin equation (\ref{eq:Lang_intr}) is
used, then $\tilde{h}(x) = 0$ and the Fokker-Planck equation for
$P_{u}(u,t)$ takes the simplest form
\begin{equation}
    \lambda^2(t)\frac{\partial}{\partial t}P_{u}(u,t) =
    \frac{\partial^{2}}{\partial u^{2}}P_{u}(u,t).
    \label{eq:FP2}
\end{equation}
Solving it, for example, by the generating function method and using
the initial condition $P(u,0) = \delta(x-x_{0})\sqrt{d(x)}$ [recall
that $x = U^{-1}(u)$], we find
\begin{equation}
    P_{u}(u,t)=\frac{1}{\sqrt{2\pi}\,\sigma(t)}\exp\!
    \bigg[-\frac{(u-u_{0})^{2}}{2\sigma^2(t)}\bigg].
    \label{eq:PDFu}
\end{equation}
Finally, substituting this result into Eq.~(\ref{eq:rel}) and taking
into account that $U(x) = \Psi(x)$, $d(x) = G^{2}(x)$ and $u =
\Psi(x)$, we again obtain Eq.~(\ref{eq:Px}).

%\clearpage

\end{document}